\documentclass[12pt]{article}

\usepackage{scicite}

\usepackage{times}

\usepackage{amsmath}
\usepackage{amsfonts}
\usepackage{amssymb}
\usepackage{graphicx}
\usepackage{endnotes}  

\newenvironment{sciabstract}{%
\begin{quote} \bf}
{\end{quote}}

\title{Definitively Identifying an\\
Inherent Limitation to Actual Cognition}

\author{Arthur Charlesworth}

\date{May 31, 2019}

\begin{document}
\maketitle 
\noindent
\verb|      |\normalsize{University of Richmond\verb|     |
DartCharlesworth@gmail.com}

\baselineskip24pt

\begin{sciabstract}
Abstract:
A century ago, discoveries of a serious kind of logical error made separately 
by several leading mathematicians led to acceptance of a sharply enhanced 
standard for rigor within what ultimately became the foundation for Computer 
Science.  
By 1931, G\"{o}del had obtained a definitive and remarkable result:
an inherent limitation to that foundation.
The resulting limitation is not applicable to actual 
human cognition, to even the smallest extent, unless both of these extremely 
brittle assumptions hold: humans are infallible reasoners and reason solely 
via formal inference rules.  Both assumptions are contradicted by empirical
data from well-known Cognitive Science experiments.  
This article investigates how a 
novel multi-part methodology recasts computability theory within Computer 
Science to obtain a definitive limitation whose application to human cognition 
avoids assumptions contradicting empirical data.  The limitation applies to 
individual humans, to finite sets of humans, and more generally to any 
real-world entity.
\end{sciabstract}


\noindent

\newpage
\section*{Introduction}
\renewcommand{\thefootnote}{\fnsymbol{footnote}}
\begin{quotation}
\noindent
Nobel-winning biologist Sydney Brenner
recently suggested
``researchers in both artificial intelligence and neuroscience might be 
getting 
overwhelmed with surface details rather than seeking the bigger 
questions underneath ...'', indicating
he was ``worried that neuro- and cognitive scientists were
being “overzealous” in these attempts''.
He recommended a ``refocus
on higher level problems instead''\cite{cass}.
\end{quotation}
A gap\footnote{Readers of this prepublication draft
are invited to email comments and questions to the author.
{\bf Each superscript number within the text 
has a corresponding NOTE in the Appendix.}
After the article gives relevant background information,
NOTE~\ref{history-note} sketches the history
mentioned in the abstract.}
exists between Cognitive Science
and mathematical results within Computer Science
whose application to humans
would require extremely brittle idealizations and would also
contradict other well-known empirical results as summarized by
Philip Johnson-Laird {\em PNAS} \cite{johnson-laird_PNAS}.
That gap is analogous to the
gap between Cognitive Science,
in which the ordering of events affects the context of human 
reasoning,
and results of Probability Theory that assume
such ordering is irrelevant
\cite{busemeyer_and_wang}.
Also analogous is the gap between Cognitive Science and
results within Mathematical Economics that assume idealizations 
about humans, a gap
bridged by Nobel-prize winning work of
Kahneman in 2002, Shiller in 2013, and 
Thaler in 2017.
A reason why bridging such gaps in transdisciplinary research 
might be difficult is that doing so requires
``tolerance for ideas
breaking with traditions'' {\em PNAS} \cite{de_bruine_and_fischoff}, p.~6; 
also see \cite{sahlin}, p.~33.

We obtain a result that helps bridge the gap 
between Cognitive Science and Computer Science.
It is important and traditional for nearly all research in Cognitive Science 
to be measured in terms of high statistical significance within the 
context of natural randomness affecting experiments.
But sometimes gaps are bridged by breaking with tradition.
The Cognitive Science result obtained has definitive certitude, 
not just high statistical significance; the related
natural randomness is addressed in our {\bf Implications} section.
To be readable by specialists from either side of the gap,
we provide background information well-known 
within Cognitive Science but not within Computer Science, and vice versa.
Scientists across a variety of fields may thus find this article accessible.

The result obtained is limitative:
that either cognition is noncomputational
or there is a metacognitive comprehensibility-related ``blind spot'', or
both.
In contrast to perception-related blindness --
such as the spot where light hits the
optic nerve, or inattentional blindness\cite{inattentional_blindness}, or
attentional blink\cite{attentional_blink}, or
 what has been recently termed
``introspection's blind spot''\cite{mapping_introspection} --
the metacognitive limitation applies to a human when perceiving
with both eyes and vigilantly attending to relevant input.
More broadly, the limitation is 
applicable to a real-world entity regardless of whether its
cognition (if any) is that of a single individual or is
based on a neural architecture.
The {\bf Implications} section discusses the application of the
limitation to any finite set of humans, as well as to that set's
understanding of
an Artificial Intelligence (AI) system that would accurately simulate
that set.
The limitation is
also very specific for each entity, as explained in {\bf Open Questions}.

Our {Main Question} is:
Could human cognition ever fully comprehend the input/output of 
human cognition?  
Our investigation of that question uses Donald Knuth's
strong computation-related 
criterion for ``understanding'', explained after the next paragraph.

Sometimes applying Kurt G{\"o}del's Second Incompleteness Theorem 
(related to Computer Science) is
suggested as a way to answer that kind of question.  We
review how requirements 
for applying that theorem to human cognition are extremely brittle and
contradict well-known empirical evidence, and why it follows that
such an application cannot succeed to even the smallest extent.  
We also investigate
how the recent COAT Theorem\cite{charlesworth_computing_reviews_2016} 
is applicable to humans and indicates a negative answer to the Main
Question.
(COAT is an acronym for Computationalism-impossible Or ``Absolute'' Truth.)

A fundamental goal of science is to define principles of 
interaction among lower-level phenomena that help explain higher-level 
phenomena; 
e.g., the classic Atomic Theory  \cite{feynman}, Ch. 1~$\S2$,
\cite{dehaene_consciousness}, p. 162.
Research investigating human cognition can seek to create and understand
simulations based on the lower-level principles
\cite{anderson2009, eliasmith2012, marcus2014}.
Forty six years ago computer scientist Donald Knuth argued
in an {\em American Scientist} 
article that 
fully achieving that goal requires expressing the relevant knowledge
algorithmically:
\begin{quotation}
\noindent
Actually a person does not {\em really} understand something until after
teaching it to a {\em computer}, i.e. expressing it as an algorithm ...
An attempt to formalize things as algorithms leads to a much deeper
understanding than if we simply try to comprehend things in the traditional
way;
\cite{knuth1996_with_page} emphasis in original,
based on \cite{knuth1973}.
\end{quotation}
Recently DeepMind's Demis Hassabis argued similarly.\endnote{ 
\label{note-on-hassabis-opinion}
In the {\em Nature} article ``Is the brain a good model for 
      machine intelligence'' \cite{hassabis_nature}, Hassabis states
      ``To advance AI [Artificial Intelligence], we need 
      to better understand the brain's workings at the algorithmic level ...
      Conversely, from a neuroscience perspective, attempting to distil
      intelligence into an algorithmic construct may prove to be the
      best path to understanding some of the enduring mysteries of
      our minds.''
      That article includes contrasting opinions by 
      Dennis Bray, Rodney Brooks, and Amnon Shashua.
      A review, by Hassabis and three of his colleagues, 
      of the effect of neuroscience on AI is
      in \cite{hassabis_neuron}.
}
Applying Knuth's criterion\endnote{Notice that Knuth's 
criterion is not satisfied by a person who develops
a Machine Learning program to
{\em teach itself} something from data.
Often the reason Machine Learning programs are developed is because
programmers realize they themselves do not fully understand something, 
such as 
exactly how they make decisions when classifying photographic images.}
to the ``understanding'' of
human cognition
requires programming and comprehending algorithms that 
accurately simulate the input/output of actual human cognition;
cf.  \cite{anderson2009,marcus2014,theil,SMM}.

A prominent attempt to show that
G{\"o}del's well-known theorem is applicable to human cognition
was made by G\"{o}del himself \cite{godel_1951}.
We review why such an attempt is unsuccessful because it fails
to satisfy 
{(I)} applicability to actual human cognition and ({II}) definitive and
robust rigor.
The article also explains how the recent COAT Theorem 
achieves applicability to actual humans and 
robust rigor by its use of a novel conceptualization to
{recast} a result of G\"{o}del, 
without requiring new results in computability theory.
After {\bf Background}, we review an intriguing
relevance of a conjecture of G\"{o}del and 
explicate two paths toward a sharper account.
The first led to {recasting} by Reinhardt of a result of G\"{o}del,
the second to the {recasting} by
the COAT Theorem mentioned above.
We end with {\bf Summary},
{\bf Implications}, {\bf Open Questions}, and {\bf Conclusions}.

\section*{Background}

As mentioned above, requirements for a definitive answer 
to the {Main Question}
include {(I)}
applicability to actual human cognition  and ({II}) definitive and robust rigor.

\subsection*{Clarifying requirement {(I)}}
It is recognized that, when actively
making decisions, human logical reasoning is not infallible.
For decades cognitive scientists have used
the prevalent fallibility of human cognition as a tool to
investigate human cognitive strategies;
e.g., \cite{newell_and_simon,martin_braine,rips,dehaene_number_sense,johnson_laird}.
Human mathematical
errors can be notoriously large; in experiments requesting magnitude 
estimation,
a third of participants gave answers roughly 30000\% different 
from the correct 
answer\cite{landy_silbert_goldin2013,landy_charlesworth_ottmar}.

In addition to the recognized lack of human infallibility,
Johnson-Laird's research team \cite{johnson_laird,johnson-laird_science}
obtained empirical evidence strongly suggesting
a ``mental models'' account of human deduction.
That team showed how such an account is predictably different from the
``mental logic'' account \cite{martin_braine,rips} 
of using the kind of logical inference rules studied within the
computability theory of Computer Science.
Johnson-Laird summarizes, in part, as follows:
\begin{quotation}
\noindent
Human reasoning is not simple, neat, and impeccable.
It is not akin to a proof in logic ... Reasoning is more a simulation of the
world fleshed out with all our relevant knowledge than a formal manipulation
of the logical skeletons of sentences
{\em PNAS}
\cite{johnson-laird_PNAS}, p. 18249.
\end{quotation}

Although human cognitive abilities are diverse \cite{levinson_diversity},
fallibility extends to those with impressive abilities:
Alan Turing, G{\"o}del, and Albert Einstein made 
mathematical errors in their individual research,\endnote{
As the main representatives for the fields of computer science,
mathematics, and physics, the {\em Time} magazine issue on 
``The Century's Greatest Minds'' (Mar. 29, 1999) chose Turing, G{\"o}del, 
and Einstein respectively.
Here are examples of errors by each in their individual research.
Turing's seminal article in computer
science \cite{turing1936} had technical errors; see
\cite{hodges}, p. 546.
A G{\"o}del biography discusses incorrect
mathematics by him \cite{dawson}, pp.~235-236.
Einstein erroneously calculated the extent to which
the sun would bend a star's light due to General Relativity;
the fame he received from the 1919 observation 
of a solar eclipse might have been diminished had it not 
been for his correction made during a delay caused by the first World 
War\cite{banesh_hoffman}, p. 133.
}
and serious concern about the fallibility of mathematicians 
is summarized in \cite{avigad_and_harrison_with_section}.
For other observations 
emphasizing the prevalence of serious fallibility of human cognition, see
\cite{charlesworth_computing_reviews_2016}, pp.~211-212.

\subsection*{Clarifying requirement {(II)}}
Neither abundant evidence for a result nor the result being
intuitively obvious is sufficient for the result to have definitive rigor.
``Definitive rigor'' requires theorems about 
mathematically-precise concepts that in
principle can be defined entirely in terms of sets
\cite{maclane_with_page}.
Daily evidence for millennia supported the intuitively obvious claim that
the Sun revolves around the Earth.
The theorems of 
Newton's Theory of Gravitation\endnote{
Newton's Theory of Gravitation
received a rigorous foundation in the 19th century\cite{boyer},
was improved by Einstein's General Relativity, and would
benefit from further understanding of
additional lower-level phenomena \cite{feynman}, Ch. 7.
}
show how a Sun-centered model provides
a simpler explanation of that high-level phenomenon
-- as well as supporting 
Kepler's assertions\endnote{
\label{note-on-kepler} 
In 1609 Kepler published a table to support his assertion 
that a planet moves in an elliptical orbit.
In 1988 very strong evidence was presented indicating
he obtained the table entries 
from his assertion itself, rather than from astronomical observations.
See \cite{donahue,wilson1994,vanderburg}.
}
about planetary motion 
--
via reduction to the physics of
precisely-defined lower-level point masses.
Another example:
Kepler in 1611, inspired by his search
for a reductionist explanation of the shape of snow crystals \cite{kepler},
conjectured that the densest ways to
stack equal-sized spheres were the intuitively sufficient
arrangements ubiquitously used for stacking cannonballs.
Despite abundant evidence, Kepler's Conjecture
was unsettled for 394 years until a recent,
celebrated theorem \cite{hales_annals, hales_forum_of_mathematics}.
Intuition also strongly supports the claim: Any non-self-intersecting curve 
within the plane that starts and ends at the same point
-- like a circle or the wobbly wall of an amoeba
pressed into the nearly planar-thin space within a microscope slide -- 
divides the entire plane into two regions exactly one of which has finite area.
Yet that claim was not definitive
until a surprisingly difficult proof by
Camille Jordan in 1893; that proof
is ``a benchmark of mathematical rigor [that] has continued to 
our day'' \cite{hales_jordan1}, p.~882, and the
resulting Jordan Curve Theorem has ``fundamental importance ... to geometry''
\cite{hales_jordan2}, p.~46.
Likewise, failure for decades to obtain a
quick (``polynomial-time'')  algorithm for solving any of a well-known set of
practical problems 
gives abundant evidence for the 
$P \neq NP$ conjecture.
Paraphrased, it states that not every problem having a corresponding
algorithm that can quickly check the correctness of a potential solution also 
has a corresponding algorithm that 
can quickly find a correct solution.
Some today might view that conjecture 
-- when it is formulated precisely --
as being intuitively obvious.
The lack of a theorem definitively settling it
(one way or the other)
has been considered the leading open question within computer science 
\cite{millennium_problems}, p. 253.

Requirement ({II}) 
also requires reasonably accurate robustness
of the real-world application of a theorem.
Fig.~1 
\begin{figure}
\verb| |
\newline
\verb|   |
{\thicklines
\begin{fbox}{
{\small  
Real-world phenomenon
} 
}
\end{fbox}
\hspace{0.2in} 
{\vector(1,0){50}}
\verb|  | 
\begin{fbox}{
{\small  
Model within 
pure math
} 
}
\end{fbox}
\newline
\verb|   |
\verb|                           |
\hspace{1.0in}
{\vector(0,-5){50}}
\newline
\verb| |
\newline
\verb|   |
\begin{fbox}{
{\small  
Application of}
{\small  
the theorem 
} 
}
\end{fbox}
\hspace{0.8in}
{\vector(-1,0){50}}
\hspace{0.1in} 
\verb|        | 
\begin{fbox}{
{\small  
Theorem from model
} 
}
\end{fbox}
} 
\newline
\verb|  |
\newline
\newline
\verb|  |
\verb|  |
{\bf Non-rigorous realm}
\verb|                | 
{\bf Rigorous realm}
\newline
\verb|  |
\verb|  |
{\bf of applied mathematics}
\verb|             |\hspace{0.03in}
{\bf of pure mathematics}
\newline
\verb|  |
\newline
\verb|  |
\newline
\verb|      |Fig.~1.  Relationship between applied and pure 
mathematics.\\
\verb|   |\\
In this article, ``definitive rigor'' refers narrowly to
theorems in pure mathematics about concepts having
mathematically-precise definitions.
That does not diminish the importance of 
the highest standards for rigor {\em within} science,
such as how methods of modern astronomy achieve much higher rigor
than the method Kepler used to support his claim of an elliptical-shaped 
orbit,
a method
whose fatal flaw was reported 379 years later; 
see Appendix NOTE~\ref{note-on-kepler}.
\end{figure}
gives a standard diagram relating applied and pure math.
For example,
the calculation of the length of
a (perfectly) circular planetary orbit 
would give reasonably accurate results, when applied to a noncircular
elliptical orbit whose eccentricity fails to be zero by
one in 10 to the 30th.
A more general example of the robustness of many
mathematical theorems is the use of differentiable functions.
Theorems about such functions typically have an underlying 
assumption that
the set of decimal numbers occurring in measurements is infinitely
divisible and ``topologically complete''.
Yet such theorems can give reasonably accurate
results when applied to the real-world, even though
the underlying assumption might not be perfectly satisfied because of 
possible discretization of space on the 
smallest scale \cite{discrete_physics_2004}.

The robust nature of hypotheses of many theorems 
contrasts sharply with the (perfectly) infallible hypothesis
within theorems about mathematical logic obtained
by G{\"o}del.
We shall use the redundant adjective ``(perfectly)'' to emphasize the
{extreme brittleness of such an infallibility hypothesis within 
such logic.
We explain the nature of that brittleness after explicating
an argument by G{\"o}del to support a 
conjecture he made.

\section*{G{\"o}del's Gibbs Conjecture} 

As we soon explain, there is intriguing relevance to the {Main Question}
of a conjecture made by G{\"o}del.
G{\"o}del asserted his conjecture in 
his 1951 Gibbs Lecture to the American Mathematical 
Society.\endnote{
Within his published Gibbs Lecture,
here is the statement of G{\"o}del's Gibbs Conjecture:
{\em 
``Either mathematics is incompletable in this sense, that its
evident axioms can never be comprised in a finite rule, that is
to say, the human mind (even within
the realm of pure mathematics) infinitely surpasses the powers of
any finite machine, or else there exist absolutely unsolvable
diophantine problems of the type specified ...''} 
\cite{godel_1951}, p. 310, italics in original.
G{\"o}del's lecture makes clear that by
a ``finite machine'' G{\"o}del means a Turing machine or its equivalent
\cite{godel_1951}, pp. 304-305.
A ``diophantine problem'' is an arithmetical problem related to the existence
of a solution in integers 
to one or more polynomial equations having integer-valued
coefficients and one or more variables. 
Earlier, G{\"o}del proved that
determining the truth value of $Con_S$ (see Fig.~3)
is equivalent to a diophantine problem.
}
G{\"o}del's Gibbs Conjecture, abbreviated GC in this article, 
is paraphrased in Fig.~2.
\begin{figure}
\begin{quotation}
\noindent
At least one of the following two assertions holds.\\
\verb|   |\\
\noindent
Assertion 1:\verb| |It is impossible for any computer program to 
accurately simulate the human mind.\\
\verb|   |\\
\noindent
Assertion 2:\verb| |There is a particular true arithmetical statement
impossible for the human mind to correctly master.\\
\verb|   |\\
\noindent
Moreover,
if Assertion 1 fails to hold --  so a computer program accurately 
simulating
the human mind could exist -- then there is a computationally constructive
way to take such a simulation program as input and produce as output
a particular arithmetical statement as described in Assertion 2
that is also {\em related to a property of the simulation program itself}. 
\end{quotation}
\verb|   |Fig.~2. Paraphrase of G{\"o}del's Gibbs Conjecture (GC).
\end{figure}

There are deep, well-known, relationships between 
the concepts of computer programs and arithmetic mentioned in Fig.~2.
Given sufficient computer memory, any
program --
in programming languages like Fortran, C, C++, Java, and 
Python -- is 
theoretically equivalent 
to a ``Turing machine'', a 
precisely-defined mathematical notion investigated by Turing in 1936
\cite{turing1936}.
The description of any Turing machine 
can be computationally encoded 
as a corresponding natural number, and
arithmetical statements are sufficient for expressing 
fundamental properties of computable functions; e.g., see \cite{hodel}, p. 387.
(The natural numbers are the nonnegative integers: 0, 1, 2, etc.)
An ``arithmetical statement'' is a statement expressed in ``formal''
Peano Arithmetic (PA), a logical system
whose language syntax is as precisely defined as that of 
a programming language, and which permits just a single variable type,
intended to mean ``natural number''.
The syntax of PA includes standard symbols for logical operations and
for zero,
addition, multiplication, successor, and less-than.
The actual possible meanings of the symbols within PA are constrained by
the axioms and the ``inference rules'' of PA.
When this article mentions the meaning of a statement of PA without
mentioning an interpretation, we intend the meaning
according to the standard interpretation of the symbols of
PA, so that the symbol + of PA is interpreted as addition of
natural numbers, and so forth.
Interpreted as such, each single statement of PA is 
exactly one of true or false, by conventional mathematics.

\subsection*{Intriguing Relevance of GC}
We now explain that 
if one were willing to ignore failure related to both 
requirements {(I)} and ({II}) mentioned in
the {\bf Introduction}, GC suggests a
``no'' answer to the {Main Question},
because of Knuth's criterion.
On the one hand if GC's Assertion 1 holds, then
it is impossible for a computer program to
exist that accurately simulates human cognition and so
(by Knuth's criterion)
it is impossible for human cognition to be fully understood.
On the other hand, if Assertion 1 fails to hold, 
then by the last paragraph of
Fig.~2
it is impossible for human cognition to correctly master an 
understanding related to such a simulation program itself.

Importantly, that last paragraph of the paraphrased GC is 
more relevant to the {Main Question}
than a claim that
human cognition cannot correctly
master all true arithmetical statements.
That weaker claim would hold
if the truth status of any specific
arithmetical assertion, regardless of its relevance to mastering any 
simulation program, should elude humans.
One candidate is an assertion that
$0$ is the $k^{th}$ digit to the right of the decimal point of $\pi$,
where $k$ is a specific huge 
number\cite{feferman_rolf_schock_with_preprint}, p. 149.
Here $k$ might be taken to be the number concisely denoted by Knuth as
$10\uparrow\uparrow\uparrow\uparrow 3$,
who suggested to scientists that in terms of magnitude it is
``so large as to be beyond human comprehension'' 
{\em Science} \cite{knuth_up_arrow}, p. 1236.

But what about requirements {(I)} and ({II})?
To see that G{\"o}del's
argument for GC 
satisfies neither, we consider his argument.

\subsection*{G{\"o}del's Argument for GC}
G\"{o}del's argument 
attempted to take his Second Incompleteness Theorem
about formal axiomatic systems, summarized in Fig.3
and apply it to humans.
G{\"o}del's rigorous theorem would fit on the right of 
Fig.~1, and his inherently nonrigorous GC on its left 
\cite{feferman_rolf_schock_with_preprint}, pp. 141-142.
\begin{figure}
G\"{o}del's 
First Incompleteness Theorem
is about a formal axiomatic system that is
computably axiomatizable.
Each computably axiomatizable formal axiomatic system $S$
has associated with it two computer programs.
One program can check the syntax of
each candidate for being a ``formal statement'', the other 
can check correctness of the inference rule applications
within each candidate for being  a ``formal proof''.
A ``formal theorem'' is any statement resulting 
from the last application of an inference rule within a formal proof.
Such a system $S$ is ``consistent'' exactly when there does not 
exist a formal statement within $S$ such that both
it and its negation have formal proofs within $S$.
Such a system $S$ is (deductively) ``incomplete'' exactly when there 
exists a formal statement within $S$ such that neither
it nor its negation has a formal proof within $S$.
The First Incompleteness Theorem (as extended by J. B.
Rosser 
\cite{hodel},
\cite{shoenfield})
states that for each computably axiomatizable 
formal axiomatic system $S$ that includes PA (Peano Arithmetic):
if $S$ is consistent, then $S$ is incomplete.
(Stated in terms of computer programs, 
a general version has a simple proof; e.g., \cite{charlesworth_godel}.)
G\"{o}del's Second Incompleteness Theorem
is a stronger result
that gives a {\em particular} example, $Con_S$, of an
arithmetical statement that is unprovable
within the system $S$.
The informal meaning of $Con_S$ (according to the standard interpretation
of the symbols of PA)
is that $S$ is consistent.
That is, {\em if $S$ is consistent}, then 
$Con_S$ is true (according to the standard interpretation)
but is unprovable within the system $S$.
\begin{center}
Fig.~3.  G\"{o}del's Incompleteness Theorems.
\end{center}
\end{figure}

G\"{o}del's 
Theorem is widely viewed as 
having a satisfiable hypothesis, since
nearly all mathematicians consider the relatively simple 
arithmetical axioms and
rules of inference of PA itself
as providing a formal system having the property
of being (perfectly) consistent.
That is altogether different from asserting that human
mathematical
reasoning itself is (perfectly) consistent.  
To see the distinction, notice analogously that on the one hand
a person might assert that the rules of the board game of chess ensure certain 
perfect properties of the game when those rules are followed, such as that no 
two pieces ever occupy the same board position after a legal move in the game.
On the other hand, 
it would be altogether different to assert that humans who play 
chess never make mistakes; cf.
\cite{charlesworth_computing_reviews_2014}, p. 466.

G\"{o}del's argument for GC assumed that, when actively making decisions,
{\em human logical reasoning is (perfectly) infallible}.\endnote{
Despite his Gibbs Conjecture,
G{\"o}del did recognize the fallibility
of actual humans, according to perhaps his leading interpreter who 
knew him personally, Hao Wang:
``...  contrary to the general 
impression, G{\"o}del affirms the 
fallibility of our mathematical intuition and
investigates the different degrees of clarity and certainty that exist
within mathematics'' \cite{wang_a_logical_journey}, p. 5.
}
We shall call that assumption the
{Infallibility Hypothesis} about human cognition.
G\"{o}del's argument also assumed such reasoning
was {\em solely via} 
a formal axiomatic system (which includes a system
similar to PA) all of whose
axioms are true and all of whose rules of inference
preserve truth.
The argument is that if such a system $S$ 
were not computably axiomatizable  
then Assertion 1 within Fig.~2
would be true, so GC would hold.
Otherwise, GC would hold for the following reason.
Since $S$ would be consistent (by the Infallibility Hypothesis), 
it would follow from the Second Incompleteness Theorem that
the arithmetical statement $Con_S$ mentioned in
Fig.~3
{\em would not be deducible within the system $S$}.
Thus, by the ``{\em solely via}'' assumption earlier in this paragraph,
Assertion 2 within Fig.~2
would be true.
(The important last paragraph of Fig.~2
follows
from the fact -- known in the 1930s -- that an appropriate
Turing machine 
can take as input the mathematical description of any computably axiomatizable
formal axiomatic system $S$ 
and then output the statement $Con_S$.)

\subsection*{Inapplicability of G\"{o}del's Theorem to Human Cognition}
The following is well-known
about the kind of system $S$ that G\"{o}del's argument
assumes underlies human cognition:
{\em if $S$ fails to be consistent,
then there is no statement within $S$ unprovable within $S$.}
For it follows from the definition of ``inconsistent'' that if $S$ were
inconsistent, there would be a statement $A$ within $S$ such that it 
and its negation each have formal proofs within $S$.
One could then choose any arithmetical statement -- such as the PA 
statement 
$Con_S$ (or even the PA statement
whose standard interpretation is the false assertion that $0 = 1$) --
and within $S$ prove
the chosen statement from $A$ and the negation of $A$ via
a well-known logical method;
e.g., see \cite{charlesworth_computing_reviews_2014}, p. 467.
To see the {extremely brittle} nature of the situation,
suppose human cognition used a system 
$S^{\prime}$ similar to the kind G{\"o}del is assuming, 
except that a single pair of inconsistent assertions
is discovered within hugely 
many assertions by $S^{\prime}$.
Not only would such a discovery falsify the Infallibility Hypothesis,
it would imply 
human cognition is 
{totally} incoherent\endnote{\label{note-on-paraconsistent} The fact that
inconsistency of a standard logical system enables 
deduction of all statements within such a system is called
the ``principle of explosion''.
The danger of an inconsistency within 
software such as databases and Artificial Intelligence programs
has stimulated research into many kinds 
of ``paraconsistent'' logics\cite{priest,avron}.
To avoid the principle of explosion's effect,
a paraconsistent logic is designed to be weaker than logical
systems relevant to G\"{o}del's Theorem.
For example 
-- using boldface English here rather than
formal logical notation and letting {\bf a}, {\bf b}, and 
{\bf c} denote natural numbers  
--
within some paraconsistent logics it is not possible
to deduce the otherwise-logical conclusion
{\bf (a equals b)}
from the combination of
{\bf (a equals c) or (a equals b)} with
{\bf (a is not equal to c)}.
}
in arithmetic,
deducing both
all false statements and all true statements!

Because G\"{o}del's argument
crucially depends on
$Con_S$ not being
deducible within $S$, his
argument
cannot succeed {to even the smallest extent}
unless the Infallibility Hypothesis holds.
Also, as indicated in the preceding paragraph,
an attempt to apply G\"{o}del's Theorem to human cognition
is highly brittle.
That {G\"{o}del's Theorem 
is not applicable to human
cognition} then follows from abundant empirical evidence contradicting
the Infallibility Hypothesis; 
recall section {\bf Clarifying requirement {(I)}}.
The problem is not just with G\"{o}del's argument, but with
{any attempt} to apply the incompleteness result 
within G\"{o}del's Theorem to human cognition.

Such an application also requires the additional assumption that
human cognition pursues mathematical truth {\em solely via}
the kind of inference rules studied within the computability area of
Computer Science.
That assumption is questionable; recall the
quotation from {\em PNAS} in 
our section {\bf Clarifying requirement {(I)}}.
Also, Cognitive Science experiments often reveal lack of
infallibility of human cognition even when the 
participants within the experiment are given
just a few short phrases to consider per trial.
There
is an infinite variety of arithmetical statements.
It follows that a
cognitive system with the computational
power and inference rules of a system like PA could,
for any prespecified natural number $k$,
deduce any selected one of the
infinitely many formal theorems whose shortest 
proof requires more than $k$ applications of the symbolic inference rules.
Here $k$ could be
Knuth's enormous number
$10\uparrow\uparrow\uparrow\uparrow 3$,
mentioned earlier.
Such processing exceeds the capacity of human brains;
e.g., see \cite{johnson-laird_PNAS} p.~18243, and
\cite{dehaene_number_sense} p.~249.

``Human cognition'' above means the cognition of a human being.
Extending the concept to encompass the collective
cognition of any particular finite set of humans 
does not undermine
the essential point made in the preceding paragraph.

We explain two paths toward a sharper account of the {Main Question}
than that given by GC.

\section*{One Path: Extending G\"{o}del's Methodology}

We will return to our focus on actual -- rather than 
idealized -- humans after this section.
In 1986, William N. Reinhardt's implication \cite{reinhardt1986_with_section}
{recast} the inclusive-or within G{\"o}del's GC.
We state Reinhardt's implication,
after giving background information also needed after this section.

Five years after G{\"o}del's 1931 incompleteness theorems 
identified a definitive limitation to infallible logic,
a theorem by Turing (stated in our next paragraph)
identified a definitive limitation to infallible computation\cite{turing1936}.
Turing's 1936 limitative
theorem used his mathematical definition of the theoretical machine now named 
after him, and he gave an argument claiming 
that definition rigorously captures the intuitive concept of computation.
That claim,
now called the Church-Turing Thesis, is
widely-accepted among mathematicians and computer scientists.
His 1936 article implies the existence of an algorithm that takes
as input the description of
any Turing machine M (with any stated input for it),
and produces as output
a statement of PA whose meaning is: M run with its stated input halts.
To ``halt'' means not to run forever.
Also, G{\"o}del showed how to computationally
encode as a natural number any syntactically defined finite concept 
(like a specific statement of PA or specific formal proof 
within PA or specific Turing machine),
with decoding also being computational.
Henceforth, 
``coding'' and ``code'' refers to
any preselected method of such coding.

Turing's theorem, called
the Unsolvability of the Halting Problem, indicates:
{\em It is
impossible for a
Turing machine P to take as input the code for
any given Turing machine M (together with input for M) and infallibly
produce corresponding
output that is the code for the correct
yes-no answer to the question ``Would M halt when run with the given input 
for M?''}.

The following result about arithmetical truth
is conventionally 
called Tarski's Undefinability Theorem:\endnote{
\label{note-on-Tarski}
G{\"o}del and Turing 
used different conceptualizations,
and there are additional
related conceptualizations
such as those of Church, Kleene, Markov, Post, and Tarski.
Often a result obtained using one such conceptualization can be obtained
using an alternative one.
Tarski made major discoveries related to the interpretation of
formal languages.
By convention, he is associated with what this article calls
``Tarski's Undefinability Theorem''; e.g.,
see \cite{hodel}, pp. 354, 390.
G{\"o}del actually discovered that result in 1930 (a few years before Tarski) 
and mentioned it in a 1931 letter he sent
to Zermelo \cite{wang_reflections}, p.~90.
Our section title
{\bf One Path: Extending G\"{o}del's Methodology}
can be viewed as referring to that
G{\"o}del discovery and also as referring to
G{\"o}del's 1933 methodology of using a symbol like Bw
\cite{godel_1933f}.}
{\em No Turing machine has the property that 
the set of single natural number inputs for which it halts 
is exactly the set of codes of true statements of PA.}
Reinhardt's implication
is in Fig.~4.
\begin{figure}
\begin{quotation}
\noindent
\verb|       |Bw~$\subseteq$~Tr~~implies that\\
\verb|       |either~~ 
$\forall e(U_e$~$\subseteq$~Tr implies $U_e$~$\subset$~Bw)\\
\verb|            |
~~or~~ Bw~$\subset$~Tr~~or both.\\
\verb|   |\\
The single symbol Bw denotes a set of codes of statements of PA,
$\subseteq$ denotes ``is a subset of'',
the single symbol Tr denotes the set 
of codes of the true statements of PA,
$\forall$ denotes ``for all'', 
$\subset$ denotes ``is a subset of, and not equal to''.
For each natural number $e$ encoding a Turing machine,
U$_e$ denotes
the set of single natural number inputs for which that Turing machine halts.
\end{quotation}
\verb|      |Fig.~4. How Reinhardt's implication 
\cite{reinhardt1986_with_section} {\em recasts}
the inclusive-or within GC.
\end{figure}
It is correct, because otherwise
its hypothesis Bw~$\subseteq$~Tr
could hold when its conclusion does not hold.
But then, using a strike through to denote the negation of a relation,
we would have
Bw~$\subseteq$~Tr,
the existence of an $e$ such that $U_e$~$\subseteq$~Tr and 
$U_e$~$\not\subset$~Bw, and Bw~$\not\subset$~Tr.
Their combination,
after producing Bw~$=$~Tr, 
would contradict $\forall e(U_e$~$\neq$~Tr), 
which is Tarski's Undefinability Theorem.

Thus the conclusion of Reinhardt's implication
holds
for {\em any} subset Bw of Tr.
Reinhardt 
wanted to interpret Bw as symbolizing
the set of codes of arithmetical statements that
are `provable by the human mind'
\cite{reinhardt1986_with_section};
henceforth, we put vague terms in single quotes.
Taking Bw to symbolize that well-known vague concept,\endnote{ 
\label{note-on-vague-provable-word}
The concept of proof, within the context of a formal axiomatic system as
discussed in Fig.~3, can be formulated in a rigorous way.
But it is notoriously vague outside such a context
(such as in Reinhardt's phrase `provable by the human mind').
Outside such a context 
perhaps the only clearly stated, desirable consequence of an arithmetical 
statement being `provable' is that the arithmetical statement would be
true (according to the standard interpretation of the
symbols of PA).
In 1933, G{\"o}del himself used a symbol like Bw 
and in the 1930s explained that such a vague `provable' concept 
was ``not accessible for mathematical treatment''
\cite{godel_1933f,godel_undecidable_diophantine_propositions}.
By that he presumably meant the concept could not be formulated in a way that
would satisfy the standard criterion for rigor \cite{maclane_with_page}.
Use of `provable' outside the context of a formal axiomatic 
system can quickly lead to incoherent paradox.
That is demonstrated (using a ``diagonal argument'')
on pp.276-277 of \cite{laforte_hayes_ford}, 
which concludes as follows:
``This paradox arises once we are willing to accept the
idea that it is actually meaningful to talk about
`the method mathematicians use to correctly decide problems'.
Like `truth', this phrase simply cannot be given an exact
meaning which mirrors in every way its informal [i.e., nonrigorous]
use, and assuming that it can leads one directly into the
kind of incoherence familiar from the liar paradox.''
The liar paradox occurs, for instance, when one realizes that the
following English phrase, which uses self-reference in
a nonrigorous way, is true if and only if it is false:
``This phrase is false''.
}
the inclusive-or in the second and third lines of his implication resembles
the inclusive-or part of GC, in the same order.
We return to Reinhardt's implication in our 
Summary.

\section*{Another Path: Extending Turing's Methodology}

Over a half-dozen 
conceptualizations related to computability have
been introduced; see Appendix NOTE~\ref{note-on-Tarski}.
We now describe an extension of Turing's conceptualization that we 
explain eliminates any need for the Infallibility Hypothesis.

\subsection*{The Conceptualization}
The following novel multipart conceptualization
assumes the widely-accepted Church-Turing Thesis.
First, notice that the
{Main Question} would be settled (with a ``no'' answer)
if the entity of `human cognition' fails to satisfy this finiteness
property:
each single input and output can be coded
using finitely many bits.
That is because an entity failing to satisfy that property\endnote{ 
\label{note-on-FiniteNumberBits} 
Our approach of showing how the {Main Question}
is easily answered if `human cognition' fails to satisfy the finiteness
property suffices for handling that case.
But here is broader background.
First, some question whether it is possible for
any algorithms to fully capture some {\em internal}
brain processes essential for the I/O of `human cognition', see opinions of 
Bray and Brooks in \cite{hassabis_nature}.
Second, certain ``neuromorphic'' chips for building some
massively parallel systems use energy-efficient analog
neuron-simulators rather than simulating neurons via
algorithms in the Turing-machine sense \cite{neuromorphic_computing}.
Third, some
view the human brain as a continuous dynamical system
\cite{vangelder,siegelmann,poznanski}.
(Accurately modeling such a system can require infinitely many bits
to represent some internal variable values.)
Fourth, that dynamical systems view has been criticized
\cite{eliasmith_2001, eliasmith_2002}.
Fifth, that view might not be incompatible with requiring
that each single input and output
be specifiable using finitely many bits;
see \cite{siegelmann_minds_and_machines}, p. 106.
(Also, for each input and output that will be most important to the COAT 
Theorem's assertion (2) -- presented in Fig.~5 -- there exists
a simple known encoding using finitely many bits.)
}
cannot be 
understood (according to Knuth's criterion), since it cannot 
be accurately 
simulated by a Turing machine \cite{hopcroft_motwani_ullman}, p.~323.
Thus we can restrict our further consideration of
real-world entities to those satisfying that finiteness property.
Next, define an ``agent'' in pure mathematics 
simply as ``a function from a subset 
of the set of natural numbers to the set of natural numbers''.
Natural numbers are sufficient for coding, not just
Turing machines and arithmetical statements, but
(by the finiteness property)
each single input or output of a real-world entity.
Consider any real-world entity making decisions about inputs, 
such as a chess player deciding on the next move, where -- like a 
chess player in an official match -- it is permitted no
outside help.\endnote{
The entity 
examined during a Turing Test is also permitted no outside help.
Turing devised that well-known test -- calling it the ``imitation
game'' -- as an entirely I/O based approach for investigating whether an
entity could `think' \cite{turing_mind}.
Whether passing such a test would be a sufficient demonstration is 
controversial\cite{searle1990}.}
The single entity could be a set of one or more real-world humans.
Let ``I/O'' abbreviate ``input/output, without regard to the
timing of inputs and outputs'', except for stipulating that 
-- as in chess -- the entity's first 
output for a given input counts as the entity's 
official output for that input
\cite{charlesworth_computing_reviews_2016}, pp. 461-464.
For that I/O concept, 
the (external) I/O 
by even a huge number of 
interacting internal algorithmic processes 
is known to be achievable 
by a single corresponding sequential algorithmic process
\cite{aaronson_physical_reality_with_page}.

The above agent concept can model
a real-world entity's (coded) I/O behavior {without}
requiring a mathematical definition of `mind' or assumptions about the
entity like any of these:
it satisfies the Infallibility Hypothesis,
it makes decisions solely via attempts to apply
formal inference rules, 
it uses a `brain'
having unchanging physical architecture while making
a decision, or it has
`conscious awareness'.
Such a real-world entity can make a mistake without being
logically required to make further mistakes, since such an entity
need not use formal inference rules like those studied within
the computability area of Computer Science.
Brittleness is not required.
That is a key advantage to using a Turing machine approach 
in our conceptualization, rather than a formal axiomatic
system approach (although the two approaches
are often viewed as being equivalent).
We also emphasize that
a real-world entity modeled by a mathematical agent
could output opposite assertions about
exactly the same Turing machine when given two separate inputs 
that are different codes for the machine;
a typical Turing machine has more than a single 
code\cite{hopcroft_motwani_ullman}, p. 369.

\subsection*{The Strategy}
Since the above conceptualization avoids requiring
the Infallibility Hypothesis, it might help answer 
the {Main Question}
if
it supports a theorem analogous to GC.
But there is an impasse, serious enough to consider abandoning the quest
for such a theorem, as explained in the rest of this paragraph.
A goal is to apply
the agent ${\cal A}$ mentioned in such a theorem to model a real-world
entity's I/O 
without requiring that the entity use
consistent reasoning.
A straightforward analogue to 
GC would be for the second half of its inclusive-or
to assert the existence of a particular 
true (coded) arithmetical statement input to ${\cal A}$ for which
${\cal A}$ could not correctly ``decide'' its truth value by outputting
(the code for) true.
But an 
agent
${\cal A}$ 
not required to be logically consistent 
might
``decide'' that each (coded) arithmetical statement 
(including each such false statement) is true.
Also, the first half of the inclusive-or in a straightforward 
analogue to
GC would claim the impossibility of accurately
simulating ${\cal A}$ computationally.
But there is a simple counterexample: a computational ${\cal A}$ 
which outputs the code for true given any single natural number input.
Thus such a straightforward strategy fails.

How to overcome that impasse?
In essence, a successful strategy for achieving the desired kind of theorem
is to ask an entity to output one of two 
opposite assertions, where (unlike ``true'' or ``false'') 
each assertion mentions the same Turing machine, using the same code 
for that machine.
As explained in {\bf The Conceptualization},
it is the first output for a given input that counts,
and one must emphasize the specific natural number code for a 
Turing machine
when in a context that avoids assuming an agent is consistent.
The following additional explanation about coding is needed before
stating the resulting theorem.
It is straightforward to define
a computational way to obtain from a number
$n$ encoding a Turing machine (and its input)
three additional numbers.
They are: the code of a statement, that we denote as $H(n)$,
of PA asserting that the Turing machine with
specific code $n$ halts; the code of the negation of $H(n)$; and
the code of the ``binding'' of the pair of codes just mentioned.
which we call the ``specific halting problem''
asking whether or not the Turing machine with specific code $n$ halts.
By conventional mathematics, 
exactly one of $H(n)$ and the negation of $H(n)$ is a true statement.
For simplicity, henceforth we often avoid explicit mention of the
coding by natural numbers, when mentioning
a specific halting problem or arithmetical statement.

\subsection*{The COAT Theorem}

Fig.~5 states the resulting recent COAT Theorem, whose
applicability to actual humans is further explained in our next section.
The above conceptualization and ``binding'' strategy enables its proof 
\cite{charlesworth_computing_reviews_2016}
to follow easily from Turing's Unsolvability of the Halting Problem,
which itself is easily proved when expressed in terms of
computer programs; e.g., \cite{charlesworth_mppr}.
Because it is achieved by {recasting} 
G\"{o}del's Second Incompleteness Theorem
using a result of Turing,
without requiring new conventional theorems in computability theory,
one might call it the ``G{\"o}del-Turing COAT Theorem''. 
\begin{figure}
\begin{quotation}
\noindent
Let $\mathbb{N}$ be the set of natural numbers.
Let ${\cal A}$ be any (mathematical) agent; i.e., any function from
a subset of $\mathbb{N}$ to $\mathbb{N}$.
At least one of the following two assertions holds.\\
\verb|  |\\
\noindent
(1)\verb| |It is impossible for any Turing machine to 
accurately simulate the I/O of ${\cal A}$.\\
\verb|  |\\
\noindent
(2)\verb| |There is a specific halting problem such that 
if it is given as input, ${\cal A}$ cannot 
output its corresponding true
arithmetical statement (thus a truth of which it is impossible --
not just highly unlikely -- for
${\cal A}$ to have full ``mastery'').\\
\verb|  |\\
\noindent
Moreover,
if (1) fails to hold -- so a Turing machine accurately simulating
${\cal A}$'s I/O
could exist -- then there is a computationally constructive way to
take the description of such a 
Turing machine as input and produce as output
a specific halting problem as described in (2) that is also
{\em related to a property of the simulating Turing machine itself}. 
\end{quotation}
\verb|         |Fig.~5.  The COAT 
Theorem\cite{charlesworth_computing_reviews_2016}.
\end{figure}
But the discussion of a 1947 Turing quote in our
upcoming {\bf Robustness} section indicates lack of awareness of
the possibility of such a theorem in the past, by 
G{\"o}del, Turing, and other mathematicians and computability specialists.
Analogous to the importance of the last paragraph of GC,
the COAT Theorem's last paragraph is related to self-comprehensibility.
The treatment in \cite{charlesworth_computing_reviews_2016}
does not require an entity to
support a decision with a proof attempt, unlike 
\cite{charlesworth_computing_reviews_2014,charlesworth_2006};
answers to questions about the conceptualization
appear in those articles and corresponding electronic supplements.
Here are three such questions.
Why is successfully proving -- alternatively, disproving --
any particular mathematical 
conjecture\endnote{
For instance, the conjecture could be at any level of 
Kleene's arithmetical hierarchy \cite{kleene_1943} p.~49, it
need not be at the low level
that contains the halting problems themselves.
That is explained in Section~9 of the on-line Supplementary Material for
\cite{charlesworth_computing_reviews_2014}.
}
according to the usual requirement for mathematical 
rigor \cite{maclane_with_page}
equivalent to successfully solving a halting problem\endnote{
In the context of the standard criterion for mathematical rigor
          \cite{maclane_with_page},
          the broad importance of halting problems is observed and
          explained in the 2014 article 
          \cite{charlesworth_computing_reviews_2014}, p.~446.
          The breadth of such problems in that context
          might not be widely-known, even among experts.
          For example, an article by other researchers in 2016
          gives a similar observation and explanation,
          and suggests that the 
          observation is contrary to the intuition of mathematicians 
          and computer scientists
          \cite{yedidia_and_aaronson}, p.~298.
}
directly
related to the conjecture?
Why are ``reflection principles'' not relevant here?
How can {\em nonrigorous} use of self-reference
be highly unreliable?
Respectively, see \cite{charlesworth_computing_reviews_2014} p. 451,
\cite{laforte_hayes_ford} pp. 276-277, and Russell's paradox
\cite{hodel} p.~15.

It follows from our discussion of an apparent impasse that the
following is not impossible for an 
agent satisfying the COAT Theorem's assertion (2):
Given as input just the true arithmetical 
statement mentioned in assertion (2)
for which the agent cannot demonstrate full mastery,
the agent would output (the code for) true.

\subsection*{Applicability of COAT Theorem}

Recall that our section {\bf The Conceptualization} justifies
restricting consideration in the rest of the article
to real-world entities satisfying the finiteness property.
As also explained in that section,
our model stipulates an entity's first output 
for a given input to be that entity's (sole) official output for 
that input.
For generality, we use the word
``entity'' rather than ``thing'', since the
latter often refers just to inanimate objects
\cite{am_heritage_dict}.
When the COAT Theorem's agent ${\cal A}$ -- 
which is merely required
to be a function from a subset of the natural numbers to the set of
natural numbers --
is applied to a real-world entity,
that entity E need not satisfy any infallibility property.

Here are illustrative examples.
First, E could be the empty set or other entity that never produces output
(a particular rock or comatose human might be examples).
For such an entity the COAT Theorem's assertion (2) holds and its assertion
(1) fails to hold because the entity's I/O can be simulated by a Turing 
machine producing no output.
Second,
E could be a particular human who incorrectly
responds ``false'' when asked to give the
truth value of $18~+~25~=~43$.
The function ${\cal A}$ corresponding to E
would map the code for that arithmetical statement
to the numerical code, say 0, for ``false''.
There need be no brittleness because there is no
restriction on any other value of that same function, if indeed
the function is defined for other numeric codes.
For example, it is possible that the same E -- and corresponding
${\cal A}$ -- outputs only correct true-false responses for many 
other individual arithmetical statements.
Third, E could be a particular (nonempty) finite set of humans
who work together to correctly give the truth value of $x~+~y~=~z$ 
for a single choice of a triple of natural numbers whose numerals have ten 
digits each, but who make an error for a single choice of a triple
when the numerals have a thousand digits each.
Fourth,
E could be a computational AI system that is probabilistic, using
pseudorandom numbers; then
assertion (1) of the COAT Theorem fails to hold;
thus assertion (2) holds and
the probability is zero that the AI system would produce 
correct output for the input mentioned in 
assertion (2) because of the impossibility of its
doing so.
As a prelude to our fifth example, we note that according to cognitive 
neuroscientist Stanislas Dehaene
``Some degree of chance may enter in a voluntary [human] choice,
but this is not an essential feature''
\cite{dehaene_consciousness}, p. 264,
and according to Chris Eliasmith ``Information processing in the brain ...
can be equally well described as continuous and noisy, {\em or}
discrete'' \cite{eliasmith_2001}, pp. 423-424, emphasis in original.
Fifth, if there is a human whose mathematical cognition depends so crucially on
genuine randomness of physical neural, synaptic, and/or neurotransmitter
processing that the resulting I/O -- for which the finiteness
property holds -- 
cannot possibly be accurately simulated using any sophisticated form of 
computational pseudorandomness, then E could be such a human and assertion (1)
would hold for E.
Sixth (mentioned again in the next section),
E could be a particular human who
gives correct answers to many mathematical problems but who, after much
deliberation,
incorrectly asserts that a certain complicated 
uncolored map lying on a flat surface
would require more than four
colors to enable adjacent map regions to have different colors.
Any of the above examples that mentions one or more humans could
be replaced by similar examples phrased in terms of other
entities, including physically distant entities
on which scientific experiments might never be feasible.

\subsection*{Robustness}

G{\"o}del's hand-written 1951 Gibbs Lecture was edited and published after 
his 1978 death,
with Stanford mathematician Solomon Feferman as Editor-in-Chief
\cite{godel_1951}.
Feferman's 
Rolf Schock Prize lecture to the
Swedish Academy of Sciences in 2003, published in
\cite{feferman_rolf_schock_with_preprint},
emphasized serious concerns about GC.
The recasting by the COAT Theorem avoids all such concerns,
including:
using a ``{\em highly idealized}
concept of the human mind'' [emphasis as in
Feferman's published lecture], 
containing an assertion that the human mind 
``infinitely surpasses'' Turing machines, 
assuming human cognition pursues mathematical truth solely by
applying inference rules, and 
assuming human cognition uses ``evident'' axioms that are not prespecified.

Concerning idealizations, Turing (the founder of AI) asserted
in 1947 that
\begin{quotation}
\noindent
... if a 
machine is expected 
to be infallible, it cannot also be intelligent.
There are several mathematical theorems which say almost
exactly that.  But {\em these theorems say nothing about how much 
intelligence may be displayed if a machine makes no pretence at 
infallibility} \cite{turing_london}, emphasis added.
\end{quotation}
The most notable additional theorems in 1947 related to G{\"o}del's Second 
Incompleteness Theorem and Turing's Unsolvability of the Halting problem
were published by Kleene in 1943 and 1936 
\cite{kleene_1943,kleene_1936}.
Like Turing, Feferman had expert familiarity with such notable theorems; see
\cite{feferman_psyche}, Section~2.3.
Neither his 2003 nor Turing's 1947 lecture suggested that any such theorem
could overcome the ``{\em highly idealized}'' or ``infallibility'' concern.
The same can be said for books on computability theory 
\cite{kleene_metamathematics,odifreddi,odifreddi_vol2,soare},
and numerous articles and books 
expressing concerns about how applying relevant theorems to actual humans 
requires assuming infallibility of the latter.\endnote{
\label{note-on-koellner}
Seven separate quotes -- by Davis, Dennett, Lutz, McDermott, Minsky, 
Barrow, and Russell and Norvig -- 
on pp.~211-212 of
\cite{charlesworth_computing_reviews_2016} question the basic
assumption of the consistency of `human cognition' that is
crucial for applying G{\"o}del's Theorem to `human cognition'; also see
Note~\ref{note-on-paraconsistent}.
Feferman's lecture expressed that same concern.
Much of the published concern about inappropriate idealizations
was stimulated by assumptions -- most notably by
J.~R.  Lucas and Roger Penrose -- 
even stronger than that basic consistency assumption.
That includes the seven quotes mentioned above, and also
in ``the concept of an idealized
human mind'' being ``problematic'' of \cite{koellner_chapter} p.~154 and
in the analysis of different ways to categorize idealizations of 
human cognition of \cite{shapiro_chapter}.
(Among the strongest of such idealization assumptions
is that an actual human 
correctly `knows' that her/his reasoning is consistent;
alternatively, correctly `knows' that 
her/his reasoning about halting problems is correct.)

Results by Reinhardt, and in 2016 by 
    Peter Koellner, show that even if --
    solely as an investigative technique -- one were to assume
    relevant idealization assumptions related to the claims of Lucas
and Penrose,
    concerns can be demonstrated {\em using logic itself} about such claims
    \cite{koellner_chapter}.
    Some claims by Lucas and Penrose are discussed briefly in
    the current author's \cite{charlesworth_2006} pp. 591-593; also see
    \cite{laforte_hayes_ford}.  The focus of the current
    article is how all idealization and infallibility assumptions 
    about `human cognition' can be avoided in a result similar to GC, so
    we only briefly mention stronger idealization assumptions.
}
In dozens of such insightful commentaries, including Feferman's about
GC, there is no discussion
about whether the significant concern in the commentaries
over infallibility assumptions about human cognition
is avoidable using any theorem 
applicable to an entity making (as Turing put it)
``no pretence at infallibility''.

That aligns with {nonexistent awareness of even the possibility of 
such a theorem}
among specialists in computability theory.
For instance, there appears to be no
consideration of a question like:
Could the need for infallibility assumptions
in a particular such claim be removed 
by replacing 
$Con_S$ with a true arithmetical
statement like the one
related to assertion (2) of the COAT Theorem?
We defer further explanation of the latter arithmetical 
statement until
the upcoming section {\bf Open Questions}.

Perhaps the most relevant generalization of
G{\"o}del's Theorem is Kleene's
Theorem VIII \cite{kleene_metamathematics}, p.~303, which Feferman
emphasized in 1995 \cite{feferman_psyche}, Section~2.3.
When stated in terms of Turing machines, Kleene's Theorem
contains an implication whose application requires the following
``soundness'' assumption:
Each conclusion reached within a given system $S$,
about the nonhalting of a Turing machine, 
 is a {\em true arithmetical statement}
supported by a correct justification, where that justification
can be checked computationally
by a single such checking program for the entire 
system $S$.\endnote{Although in general soundness is an even
stronger assumption
than consistency, it is known (see
\cite{feferman_psyche} Section~2.8)
that
the consistency assumption on the $S$ mentioned in Fig.~3 is equivalent to the
assumption that such an $S$ satisfies soundness
for its results about the nonhalting of Turing machines.
Also, each statement asserting the nonhalting of a Turing machine is
expressible in PA's formal language, hence also in the formal language of
the $S$ mentioned in Fig.~3.}
The $S$ in Kleene's Theorem need not include PA,
need only relate to the nonhalting of Turing machines, and the checking 
program need not relate to the specific details of PA.
But the
$S$ of Kleene's Theorem lacks applicability to 
`human cognition'.
Such applicability would require
the assumption that the
the I/O of `human cognition' is correct
whenever asserting the truth of what mathematicians call
a $\Pi_1$ statement, which is a
claim about the nonhalting of a particular Turing machine.
That infallibility assumption -- ''soundness for $\Pi_1$ statements'' --
conflicts with the empirical evidence about the fallibility of
`human cognition'; recall section {\bf Clarifying requirement {(I)}}.
Such an assumption contrasts
with the lack of any infallibility assumption on the agent
mentioned in the COAT Theorem.  

It is well-known that correctly deciding whether or not a particular
$\Pi_1$ statement is true can be highly challenging.
One such statement
is the unsolved Riemann Hypothesis, conjectured in 1859, which
``has been the Holy Grail of mathematics for a 
century and a half'' \cite{borwein_et_al}.
Settling it earns a million-dollar award from the Clay Mathematics
Institute \cite{millennium_problems}.
Others include the unsolved Goldbach Conjecture -- over 275 years old -- as well
as Fermat's Last Theorem (FLT) and the Four Color Theorem (4CT)
(``four colors always suffice'',
related to the sixth example immediately preceding
the current section).
That each of those three
is equivalent to a corresponding 
$\Pi_1$ statement is easily 
seen by considering, for each, 
whether a corresponding program that exhaustively
seeks a counterexample 
would fail to halt\cite{charlesworth_computing_reviews_2014} p.~443.
Turing himself ran programs that systematically sought a 
counterexample to the Riemann Hypothesis \cite{hodges}, pp.~408, 409, 411.
Turing presumably knew that the truth of that conjecture is equivalent
to the nonhalting of such a corresponding program; also see
\cite{yedidia_and_aaronson}, Section~5.
FLT and the 4CT were perhaps the most widely-publicized new theorems of the 
past fifty years.
Their full proofs greatly exceed their published proofs
in \cite{wiles,appel_and_haken_ill_j_math,
appel_and_haken_and_koch_ill_j_math}; see related observation in
\cite{mclarty}, p. 359.
The assertion of FLT
-- based on the claimed existence of
a proof too long to fit into the margin 
of a book -- was made in
1637, yet its truth was not accepted until a proof by
Andrew Wiles 358 years later \cite{singh}.
The 4CT was asserted in 1879 and 1880
in separate journal articles (by different mathematicians),
whose corresponding published proofs were each separately
shown more than a decade after publication to be 
incorrect \cite{update_on_four_color_thm}.
There is now a (computer-assisted) proof of the 4CT satisfying 
the kind of computably-checkable criterion that Kleene's Theorem 
requires \cite{gonthier}.
That contrasts with
the current situation regarding the proof of FLT; 
see the discussion of that proof, and of PA
and Zermelo Fraenkel set theory (ZF) in
\cite{mclarty}.
Books like \cite{hodel,shoenfield} give solid introductions to the
important example ZF of the $S$ described in Fig.~3;
brief information about ZF is in the present author's 
\cite{charlesworth_computing_reviews_2014}, pp. 445-447.

The examples given above are just four of the 
infinitely many conjectures of the kind Kleene's Theorem is
about.
The COAT Theorem mentions just a
single such conjecture, where the conjecture
is related to the agent mentioned in that theorem; further
clarification is in our {\bf Open Questions} section.
When applying the COAT Theorem to a real-world entity, 
any other possible attempts by the entity to respond to inputs are
irrelevant.\endnote{Also, it is well-known in computability theory that 
computable checks are only important when there are multiple
decisions (or proofs) to check.
A computable check on the correctness of any single correct decision is 
trivially (and correctly, but vacuously) achieved 
by a computer program that ignores its input and prints ``Correct!''.}

\section*{Summary}

Reinhardt's {recasting} has a rigorous proof 
and resembles the 
nonrigorous inclusive-or part of GC.
But, when interpreted in the real-world (recall Fig.~1),
the interpretation  `provable by the human mind' 
\cite{reinhardt1986_with_section} 
for its symbol Bw
is recognized as being fatally flawed;
see Appendix NOTE~\ref{note-on-vague-provable-word}.
Thus Reinhardt's implication is not successful in the context of this
article.

The {recasting} by the COAT Theorem 
satisfies requirement {(I)} mentioned 
in the {\bf Introduction}.
That is because, as illustrated in the section 
{\bf Applicability of COAT Theorem}, its
agent concept is not required to
satisfy the Infallibility Hypothesis or any correctness property,
and is not required to use the kind of inference
rules studied within the computability area of Computer Science.
(A general result, it is also compatible with such 
unsupported assumptions about `human cognition', and in that
context it makes rigorous the version
of G{\"o}del's nonrigorous Gibbs Conjecture
paraphrased as GC in Fig.~2.)
The COAT Theorem also satisfies 
requirement ({II}), because
its statement uses conventional mathematical concepts
(in a novel way), its proof
uses known mathematical results (in a 
novel way), and its
robustness is illustrated in the 
{\bf Applicability of COAT Theorem} section.

The COAT Theorem bridges a gap between Cognitive
Science and Computer Science.
Mathematical logic books that laid a foundation for computability
explain how the concept of rigor within mathematics
was sharply enhanced
in reaction to discoveries in the early
twentieth century about human fallibility, particularly
mistakes in logic and set theory
by leading mathematicians.
Chapter III of Kleene's book 
\cite{kleene_metamathematics}
lucidly describes that history.
But the rest of his book (like other such books)
is focused on idealized mathematics, without evident
concern for providing a rigorous result that could be {\em applied}
to not-necessarily infallible 
humans.\endnote{\label{history-note}
Here is a sketch of relevant
history.
In 1854, George Boole 
published
a logic book {\em The Laws of Thought}, showing how 
symbolic manipulations similar to those of current high-school algebra 
could be used within logic.
In 1879, Gottlob
Frege published another logic book, part of whose German title translated
to English could be paraphrased as 
A Formal Language for Pure Thought Modeled on Arithmetic.
Although both titles mention [human] thought, those authors
lacked knowledge of the more recent cognitive science research mentioned in 
our section {\bf Clarifying requirement {(I)}}.
Bertrand Russell discovered a devastating inconsistency (not a mere typo)
in the 1903 preprint of the second edition of that Frege book.
That inconsistency, now known as Russell's Paradox, is
mentioned shortly before our 
section {\bf Applicability of COAT Theorem}.
At least four other leading mathematicians in the early twentieth-century
also separately proposed serious systems of logic having
devastating inconsistency; see the Davis quote on pp.~211-212 of 
\cite{charlesworth_computing_reviews_2016}.
That is why our
abstract begins with
``A century ago, discoveries of a serious kind of logical error made 
separately by several leading mathematicians led to acceptance of a 
sharply enhanced standard for rigor ...''.
That ``enhanced standard for rigor'', 
based in set theory \cite{maclane_with_page},
would be used on the right side of Fig.~1.
The relation between applied math and pure math
depicted in that figure can be viewed as a highly-successful revolution 
in the early twentieth century, similar in some ways -- but less 
widely-known -- than revolutions within the physical sciences
in the late nineteenth and early twentieth century \cite{quinn}.
In the decades after Frege, 
mathematicians who carried out research in logic -- which 
led to the modern foundation for both mathematics and computer science --
had two primary interests.
First,
protecting the field of mathematics 
from collapse caused by discovery at some time of an inconsistency within 
the field.
Second,
proving properties related to systems like the $S$ in Fig.~3,
including properties of various ways of interpreting the symbols in such a 
system.
Such mathematicians
seem not to have been concerned about obtaining a rigorous result
that could be applied to 
actual human cognition
(including the cognition of the leading mathematicians
who had themselves committed serious errors of inconsistent reasoning).
Also see Note~\ref{note-on-paraconsistent}.

}
(Recall the analogous distinction made between chess and chess 
players, in our section {\bf Intriguing Relevance of GC}.)
Had he such concern -- and with obtaining a 
perspicuous result that {scientists could apply to 
`human cognition' without having to make highly questionable assumptions
about humans} --
he should have aimed for a theorem
that achieves such applicability.
(Our {\bf Robustness} section discusses the
nonexistent awareness of even the possibility of such a result
among specialists in computability theory.)
Doing so would also have required him to answer relevant
questions.
(See the questions answered in
\cite{charlesworth_computing_reviews_2014}, including its section
{\bf Applying the Comprehensibility Theorem to Real-World Agents}
and its online supplementary material.)

When expressing concern about infallibility assumptions on
human cognition in claims like GC, mathematicians and computability 
specialists -- including Turing and Feferman --
have been correct in not suggesting any (previous) theorem 
provides an analogue to GC that avoids dependence on such assumptions.
Conventional computability theorems lack the 
applicability of the conceptualization supporting the
{recasting} by the COAT Theorem,
and the rigorous version of GC obtained via its ``binding'' strategy
and ``mastery'' concept.

\section*{Implications}

Concerning the COAT Theorem's assertion (1), notice that
if everything within an entity (see Appendix
NOTE~\ref{note-on-FiniteNumberBits}),
including the timings of internal events, 
can be accurately simulated by a Turing machine, so can its 
I/O.
It follows -- using the
contrapositive\endnote{
The implication ``p implies q'' can be written ``if p, then q'';
that implication is logically equivalent to its 
{\em contrapositive} ``not q implies not p''. 
The following two implications are not logically equivalent:
``p implies q'' and its {\em converse} ``q implies p''.
}
of the 
just-stated implication --
that if it is impossible for any Turing machine to 
accurately simulate the entity's I/O,
then {it is impossible for any Turing machine
to accurately simulate the entity itself}.\endnote{
\label{note-on-HumanComputerHybrid}
At this point we can make
these observations:
First, the COAT Theorem can be applied to an agent that is the
(external) I/O function of
a real-world entity composed of
one or more humans and one or more computer programs.
The computer programs might be used, for instance,
to handle coding related to natural numbers, and to assist
with mathematical proofs \cite{avigad_and_harrison_with_section} 
when such proofs are possible, to help
ensure many of the entity's output assertions are correct.
Second, if assertion (1) holds for such an agent,
then the `human cognition' part of the entity
could not be accurately simulated 
computationally, since the other part(s) of the entity could be
accurately simulated computationally.
Third, if assertion (1) does not hold, then
assertion (2) holds, so
even using the computer programs (and correct
hardware) `human cognition' 
cannot give the correct output for the input
mentioned in that assertion.
}
(The converse need not hold; analogously a human chess-player could use
a memorized algorithm, a fact independent of
whether `human cognition' can be accurately simulated computationally.)

Due to the universal nature of the {Main Question},
a single situation can justify a ``no'' answer.
To see how the COAT Theorem indicates such an answer,
let E be any real-world entity 
and let the
COAT Theorem's agent ${\cal A}$ be E's I/O function.
On the one hand, if
assertion (1) holds, then by the preceding paragraph (and Knuth's 
criterion) it is impossible for E to be fully comprehensible
by any entity.
Among other possibilities, relevant here is a possible assertion that
{genuine randomness within neural circuitry} -- that is
impossible to simulate even using
extremely sophisticated computational pseudorandomness -- {is
essential for human cognition}.
Some, such as Dehaene and Eliasmith, would question that
assertion, as mentioned in the discussion preceding
the fifth example in section {\bf Applicability of COAT 
Theorem}. 
On the other hand, if assertion (1) does not hold, then 
E is not fully comprehensible by E
because, by the COAT Theorem's last paragraph,
there is a specific true arithmetical statement A {related to} E
for which E cannot have full mastery.
In either case E cannot fully comprehend E \endnote{
The following two comments are relevant here.
First, whenever that particular application of E
is made, there is of course an implicit assumption that at least one 
actual human exists; otherwise the Main Question is unimportant.
Second, for an {\em interesting} treatment of the Main Question,
the entity E1 that is to be understood should have a high level of
`human cognition'; as a simple extreme contrast, it might be easy to
build and understand a computer simulation of the 
I/O of a human if that human were comatose.
Likewise the entity E2 attempting to do the understanding should have
a high level of `human cognition'; as a simple extreme contrast, it might
be unimportant to say that a comatose human might not be able to demonstrate
an understanding.
Also, it is simplifying to focus on the case in which entity
E1 and entity E2 are the same entity, E.
Doing so is sufficient, since
a ``no'' answer to the Main Question 
when E is used twice in that way 
implies that
`human cognition' cannot fully comprehend `human cognition'.
}
There is no need to ignore
Feferman's concerns about GC
or abundant
evidence contradicting the Infallibility Hypothesis.

Here are three additional observations about a situation when
the COAT Theorem's assertion (2) holds and the agent in the theorem
is the I/O function of a real-world entity E.
First, assertion (2) indicates E has a ``blind spot'' for
recognizing a specific arithmetical truth related to E itself, thus
a {metacognitive blind spot}
more fundamental than the 
{\em perception-related} blindness of humans mentioned 
in the {\bf Introduction}.
Cognitive Science research on metacognition was stimulated by
a 1970 {\em Science} article 
about the first nonhuman animal passing
the mirror self-recognition test
\cite{gallup}, and
now extends beyond investigating self-recognition.\endnote{
Research on metacognition 
includes investigating an entity's 
own confidence in its answers
\cite{dehaene_consciousness} p. 244ff, even
when such confidence requires no separate mechanism.
There is experimental evidence that neurons in the parietal cortex
of rhesus monkeys
encode confidence level as an {\em integral part} of decision making 
\cite{kiani_shadlen}.
We also note that a 1990 mathematical technique
by J. S. Bridle can be viewed as encoding
confidence level as an integral part of
the decision making of artificial neural
networks, by using ``softmax'' (also called ``the normalized
exponential'') to construct the activation function for each 
output layer unit \cite{bishop}.
}
Second, phrasing the first observation somewhat differently:
the truth of the arithmetical statement in assertion (2) 
is an example of an {observer-independent} truth, if one 
considers the observer to be E.
Third -- related to neuroscience when E is human --
if E's output for the input mentioned in assertion (2)
were fully
determined by some internal state of E,
such a state could not produce an output that is correct; 
that would mean the comprehensibility 
limitation given by assertion (2) would not just be 
a limitation on the external behavior of E.\endnote{
In 2005, the year before he died, 
the logician
Torkel Franz\'{e}n published the generally insightful book
{\em G{\"o}del's Theorem: An Incomplete Guide to Its Use and Abuse}
\cite{franzen}.
It has two passages we realize should have been written differently.
First,
an argument on its p.~125 applies self-reference in a nonrigorous
way to intuition about human behavior,
without reminding the reader that on its p.~86 the book had pointed out 
the highly unreliable nature of arguments using self-reference in
a nonrigorous way.
(See the related discussion, shortly before our
{\bf Applicability of COAT Theorem} section,
and also see the end of Appendix 
NOTE~\ref{note-on-vague-provable-word}.
Also recall from our
{\bf Clarifying requirement (II)} section that strong intuition 
is insufficient for achieving a definitive result.)
Second, the middle of its p.~126 implicitly mentions 
both the assumption of the Infallibility Hypothesis and 
the assumption that humans reason solely via formal inference rules,
without pointing out that, within empirical cognitive science, the first 
assumption is well-known to be false and there is strong empirical
evidence questioning the second assumption.
That passage is further evidence of the gap between
Cognitive Science and Computer Science mentioned 
in the {\bf Introduction}
and explicated in our {\bf Robustness} section.
}

Here is a summary, when the COAT Theorem is applied to humans.
\begin{quotation}
\noindent
Summary: Attempts at computationally-reductionist accounts of
`human cognition' have an inherent specific limitation in
which one or both of the following hold: 
such simulation software must always fall short of full accuracy;
`human cognition' could not fully master a
comprehension of such simulation software,
regardless of how such software was obtained\endnote{
Here, ``regardless of how such software was obtained'' means
even if the software were (somehow!) obtained from an
infallible intelligence.
In recent decades, the discipline of
Software Engineering has been substantially revised, 
to better accommodate the repeated fallibility of human
software developers that is widely 
recognized
by such developers themselves; e.g., see \cite{schach}.
} 
and 
--
by the Church-Turing Thesis --
regardless of advances in programming languages and software 
engineering.
\end{quotation}
The limitation does not just apply to separate individuals.
For example, the limitation is applicable to an entity consisting of 
any particular finite set -- say H -- of humans, 
which could include specialists in Cognitive Science and Computer Science
and other disciplines,
with the restriction that H (like a set of 
chess players on a single chess team) is not permitted external help.
(Also, in such a set's attempt to
master an understanding of software that accurately
simulates the set's collective cognition -- if such software should exist --
a computer system within the set could help the humans by carrying out
lengthy and tedious details;
see Appendix
Note~\ref{note-on-HumanComputerHybrid}.)
It might not be surprising if H could not 
understand a computational AI system {\em superior} to H's cognition.
But note that it follows from the Summary that
H could not master 
an understanding of an AI system that 
accurately simulates the cognition of H itself.\endnote{It
is recognized that specialists today lack satisfactory 
understanding of systems substantially weaker than a highly
nontrivial fully-human level AI.
In 2018, {\em Science} 
 	journal's Web page reported
 	strong support at a recent AI research conference, for a speaker's 
 	comparison of Machine Learning research with alchemy\cite{hutson}.
 	In a 2011 panel discussion at MIT, 
        Noam Chomsky ``derided researchers in 
 	machine learning who use purely statistical methods to produce behavior 
 	that mimics something in the world'' but do not ``try to understand 
 	the meaning of that behavior'', asserting he does not 
 	``know of anything like it in the history of science''\cite{cass}.
 	Also see \cite{charlesworth_neural_networking}.}

The COAT Theorem, like $2+2=4$, is on the rigorous pure mathematics side
of Fig.~1.
The above Summary is on the applied mathematics side, because
the Summary lacks full precision due to its use of such concepts
as `human cognition'.
The same lack of full precision occurs
with other applications of pure mathematics such as
``2 golfers plus 2 golfers equals 4 golfers'' (since `golfer' lacks full
precision).
That assertion about a number of golfers, 
like the Summary, is {\em intended to be taken 
literally} rather than metaphorically.

Occasionally one might see a suggestion that 
G{\"o}del's Second Incompleteness Theorem provides an
adequate applicable foundation 
(in the context of interest to this article).
But such a suggestion is known to be erroneous.
Douglas Hofstadter's 1980 Pulitzer Prize winning book
{\em G{\"o}del, Escher, Bach} introduced many readers to
G{\"o}del's Theorem.
Just before discussing whether 
G{\"o}del's Theorem has relevance to 
minds and brains, that author writes
``it can have suggestive value 
to translate G{\"o}del's Theorem into other domains, provided
one specifies in advance that the translations are
metaphorical and are {\em not intended to be taken 
literally}''\cite{hofstadter_geb_with_page}, emphasis added.
See our section
{\bf Inapplicability of G{\"o}del's Theorem to Human Cognition}.

\section*{Open Questions}
As a prelude to mentioning open questions, we give brief relevant
history and also compare GC with its {recasting}
by the COAT Theorem.
Merely for
expository simplicity, to avoid repetition
throughout this explanation we assume -- as do most 
mathematicians -- the consistency of PA.
G{\"o}del's Second Incompleteness Theorem was successful in showing
that a goal (proving the consistency of mathematics)
announced in the 1920s by the prominent mathematician 
David Hilbert could not be achieved using a method similar to 
the highly reliable one prescribed by Hilbert.  
Assuming the Church-Turing Thesis,
Turing's Unsolvability of the Halting Problem was also successful, in showing
that another goal announced in the 1920s by Hilbert and his colleague 
Wilhelm Ackermann could not be achieved (that of
finding a computational way to correctly decide
whether or not any given statement within a formal system like PA
is a formal theorem).
Although both results were fully successful, when initially obtained they 
both might have seemed narrowly limited to self-reference decisions 
about what can be viewed as halting problems (our next paragraph
explains how $Con_S$ can be viewed as such a problem).
But there have been surprising generalities discovered related to both.
It suffices here to focus just on Turing's halting problem result.
One surprise: Rice's Theorem \cite{rice}
identifies a huge variety of algorithmically unsolvable 
(and non self-reference) decision problems; it is proved
using the
fact that programs (in any standard programming language,
including that of Turing machines) can be written that contain infinite loops
\cite{hopcroft_motwani_ullman} p.~389.
Another surprise: theorems in 2002 prove the existence of important
algorithmically unsolvable (non self-reference)
decisions about programs whose programming language 
syntax makes infinite loops {\em impossible}
to write \cite{charlesworth_toplas2}.
Other historical surprises include the infinitely-many different levels of 
the arithmetical hierarchy \cite{kleene_1943} p.~49, and
the extreme complexity of the uncountably large 
partially-ordered set 
of Turing degrees \cite{odifreddi_vol2}.
See descriptions of additional
algorithmically unsolvable problems
in \cite{davis_unsolvable,poonen_unsolvable} and in
computability books; e.g., \cite{odifreddi,soare}.

Now we compare the nonrigorous GC with the COAT Theorem.
Assertion~2 of GC refers to an arithmetical statement.
In G{\"o}del's argument, that statement 
is $Con_S$, where $S$ is the 
(perfectly) consistent formal
system presumed in
his argument to
underlie `human cognition'.
Equivalently, {\em $Con_S$
states that the following
algorithm would fail to halt: an exhaustive search for
two blatantly contradictory formal theorems 
(i.e., a formal theorem and also its formal negation)
within $S$}.
Because of its supporting
conceptualization, including
its ``binding'' strategy and ``mastery'' concept,
the COAT Theorem's last paragraph gives the
following analogue to GC's last paragraph, 
without needing any correctness assumption about 
an entity corresponding to
${\cal A}$.
If its assertion (1) is false, then
{\em the true arithmetical statement mentioned in its assertion (2)
can be constructed to state exactly one of the following 
about a specific algorithmically-obtainable variant of the software
simulating ${\cal A}$ (with corresponding specific input):
it halts, it does not halt.}
[Correctness of the preceding sentence is easily seen by specialists
in computability theory; e.g., there is an easy proof
of Lemma~7.15 on p.~601 of \cite{charlesworth_2006},
which is restated more simply on p.~455 of 
\cite{charlesworth_computing_reviews_2014},
and used near the end of the COAT Theorem's proof atop p.~220 of
\cite{charlesworth_computing_reviews_2016}.]

The true arithmetical statements mentioned in both Assertion~(2) of GC and 
assertion (2) of the COAT Theorem are very specific.
Although the second paragraph of our {\bf Implications} section shows
the COAT Theorem is successful in indicating a ``no'' answer to 
the Main Question, there
are open questions about the breadth of the phenomenon it identifies,
just as there would have been such questions about
G{\"o}del's Theorem and Turing's Unsolvability of the Halting Problem when
they first appeared.
One open question is:
What additional theorems -- like the COAT Theorem --
can be discovered that are 
about an agent, are applicable to real-world entities,
and permit avoiding all infallibility requirements on the
agent?
Another open question: What theorems can be discovered,
if one is willing to place some limited
correctness requirements on the agent?
One answer to that question is the Comprehensibility Theorem
\cite{charlesworth_computing_reviews_2014},
p.460.

\section*{Conclusions}

\bigskip
Rather than a definitive limitation like G{\"o}del's 
Theorem about infallible logic or a definitive limitation like 
Turing's Unsolvability of the Halting Problem about
infallible computation,
the COAT Theorem is a definitive limitation about 
not-necessarily infallible and not-necessarily purely-deductive
comprehensibility.
Applicable to any real-world entity, it was
obtained by recasting computability theory within Computer Science.
The {\bf Implications} section explains how the limitation is related
to AI, to Cognitive Science, and to neuroscience.

Identifying limitations applicable in the natural world
clarifies science's boundaries.
Significant such limitations include
Heisenberg's Uncertainty Principle and 
the speed-of-light limit given by Special Relativity.
But, just as that speed limit never diminished support for faster
particle accelerators,
the COAT Theorem does not diminish the importance of continuing
scientific
progress in cognitive neuroscience and
simulations of the mind and brain 
\cite{anderson2009,marcus2014,theil,SMM}.
An open-ended question is whether additional applicable
and definitive results can be obtained
about not-necessarily infallible human cognition that is conceptually
above the level of neural interactions.
\bigskip

\newpage

\newpage
\begin{center}
{\huge
{\bf Appendix}
}
\end{center}
\theendnotes

\newpage
\bibliography{cognition}
\bibliographystyle{ScienceAdvances}

\section*{Acknowledgments}
The author is grateful to Jonathan D. Charlesworth,
James A. Davis, Gary R. Greenfield, Peter Koellner,
Donald W. Loveland, William T. Ross, and
Jack E. Singal for helping to improve the exposition.
This work is dedicated to the author's Duke University graduate
school professors Richard E. Hodel and the late Joseph R. Shoenfield.
\end{document}